\documentclass{optica-article}

\journal{opticajournal} 

\articletype{Research Article}

\usepackage{booktabs}
\usepackage{multirow}
\usepackage{lineno}
\usepackage{tabularx}

\begin{document}

\title{CSPR-Net: Self-supervised Curved Surface Projection Rectification Network for Geometric Distortion Correction in Non-planar Projections}

\author{Kejin Peng\authormark{1}, Jia Wei\authormark{2,*} and Xiang Hao\authormark{1,*}}

\address{\authormark{1}College of Optical Science and Engineering, Zhejiang University, Hangzhou 310027, China\\
         \authormark{2}Department of Urology, Children's Hospital, Zhejiang University School of Medicine, Hangzhou 310052, China}

\email{\authormark{*}haox@zju.edu.cn;dr.weijia@zju.edu.cn} 


\begin{abstract*} 
Projecting images onto non-planar surfaces inevitably introduces geometric distortions that degrade visual quality.
Traditional correction methods often require tedious manual calibration or structured light sequences to establish pixel-wise correspondences.
In this paper, we develop the Curved Surface Projection Rectification Network (\textbf{CSPR-Net}), a self-supervised deep learning framework for automated distortion correction.
Our approach employs dual coordinate-based neural networks to learn the bi-directional mapping between the projector and camera spaces.
By enforcing a robust cycle-consistency constraint, \textbf{CSPR-Net} autonomously resolves complex geometric transformations without requiring ground-truth deformation fields.
Furthermore, a gradient-based loss function is introduced to mitigate the impact of complex ambient light interference and accurately capture high-frequency geometric variations.
Quantitative evaluations in physical experimental scenarios demonstrate that \textbf{CSPR-Net} achieves a 20.7\% improvement in end-to-end fidelity (SSIM) and outperforms the polynomial baseline by 3.8\% and 5.4\% in forward and inverse mapping in terms of SSIM respectively, effectively generating high-precision pre-warped images for seamless projection.

\end{abstract*}

\section{Introduction}
Geometric distortion correction is a critical prerequisite for high-fidelity spatial augmented reality and projection mapping \cite{grundhofer2018recent,bimber2005spatial,park2016defocus,ibrahim2024spatially,tehrani2019automated}. When the projection surface is non-planar, such as a cylindrical or arbitrarily curved manifold, the resulting image suffers from severe warping that degrades the viewer's visual experience.

Traditional correction strategies generally follow two distinct paradigms: feature-based mapping and geometric reconstruction.
Feature-based methods typically utilize structured light sequences, such as checkerboards or Gray code patterns, to establish pixel-wise correspondences between the projector and camera planes \cite{moreno2012simple,ibrahim2024spatially,jordan2010projector,kio2016distortion}.
To derive a continuous transformation from these discrete correspondences, global parametric models like polynomial fitting \cite{liu2015accurate, fitzgibbon2001simultaneous} are frequently employed for their computational tractability.
However, such parametric approaches inherently struggle to model high-frequency local distortions on complex surface topologies due to their limited degrees of freedom.
Furthermore, the prerequisite feature detection stage is highly sensitive to ambient illumination and surface reflectance properties \cite{xu2007robust}.
Such environmental interference often leads to the failure of automated detection algorithms, necessitating laborious manual feature labeling that introduces subjective errors.
Alternatively, reconstruction-based approaches recover the physical topography of the surface using 3D scanners or binocular vision, subsequently deriving coordinate transformations via physical ray-tracing models \cite{manevarthe2018geometric,ibrahim2024spatially}. However, these methods require specialized, high-cost hardware and involve exhaustive, multi-device calibration procedures that limit their flexibility in dynamic environments.

Recently, deep learning-based approaches have emerged as a promising alternative \cite{lecun2015deep,barbastathis2019use}. 
Supervised learning methods typically utilize Convolutional Neural Networks (CNNs) to directly regress distortion fields by training on large-scale datasets with ground-truth deformation maps \cite{krizhevsky2012imagenet,li2019blind,jin2023perspective}. 
However, acquiring such pixel-perfect ground truth in physical environments is extremely labor-intensive and often requires complex auxiliary setups, restricting their scalability.

To circumvent these limitations, we propose Curved Surface Projection Rectification Network (\textbf{CSPR-Net}).
Rather than relying on explicit geometric reconstruction or fragile feature extraction, our approach utilizes an implicit coordinate representation to learn the end-to-end transformation \cite{wang2024implicit,huang2019compennet,isola2017image,zhu2017unpaired}. The key contributions of this work include:

\begin{enumerate}
    \item \textbf{Dual-path Neural Architecture:} A bijective mapping system that learns forward and backward geometric transformations simultaneously, ensuring spatial consistency through cycle-consistency constraints.
    
    \item \textbf{Self-supervised Training Paradigm:} A fully self-supervised strategy that circumvents the need for ground-truth deformation fields or manual annotations, enabling the autonomous learning of complex geometric mappings directly from raw projection-capture pairs.
    
    \item \textbf{Robust Gradient-based Optimization:} The introduction of a gradient-consistency loss that effectively mitigates photometric discrepancies induced by non-uniform illumination and surface reflectance variations, significantly enhancing the model's structural fidelity and environmental robustness.
\end{enumerate}

\section{Self-supervised Curved Surface Projection Rectification Network}

\subsection{Problem Formulation}
Let $\Omega_p \subset \mathbb{R}^2$ and $\Omega_c \subset \mathbb{R}^2$ denote the discrete coordinate domains of the projector and the camera, respectively. An original reference pattern $I_p(\mathbf{u}_p)$ defined on $\mathbf{u}_p \in \Omega_p$ is projected onto a non-planar surface, resulting in a distorted image $I_c(\mathbf{u}_c)$ captured at $\mathbf{u}_c \in \Omega_c$. 

Our objective is to resolve the non-linear geometric transformations $\mathcal{F}: \Omega_p \to \Omega_c$ (forward mapping) and $\mathcal{G}: \Omega_c \to \Omega_p$ (backward mapping). The forward mapping $\mathcal{F}$ models the projection-to-camera distortion, while $\mathcal{G}$ facilitates geometric reconstruction such that the reconstructed image $\hat{I}_p(\mathbf{u}_p) = I_c(\mathcal{F}(\mathbf{u}_p))$ aligns with the original pattern $I_p$. Furthermore, for any target content $I_{target}$ intended for display, a pre-warped image is generated as $I_{pre}(\mathbf{u}_p) = I_{target}(\mathcal{F}^{-1}(\mathbf{u}_c))$, which compensates for the surface curvature to produce an undistorted display from the observer's perspective.

\subsection{Neural Network Architecture}

\begin{figure}[!ht]
    \centering
    \includegraphics[width=1\linewidth]{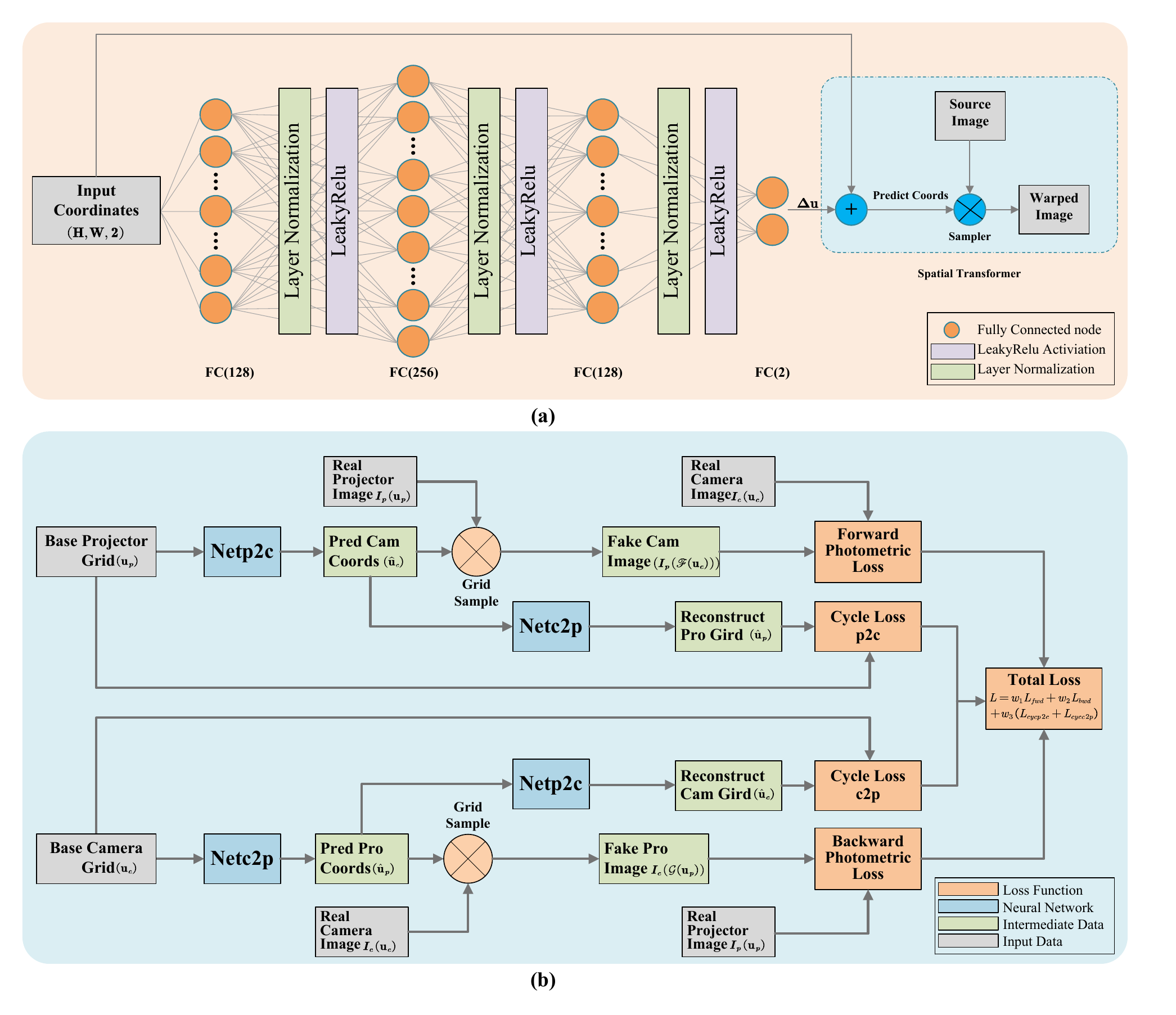}
    \caption{Overview of the proposed \textbf{CSPR-Net}. (a) Detailed architecture of \textbf{CSPR-Net}. The model utilizes a 4-layer fully connected MLP with Layer Normalization and LeakyReLU activations to regress coordinate displacements. A differentiable Spatial Transformer is integrated for end-to-end image warping; (b) The self-supervised training pipeline. The pipeline enforces spatial consistency between the projector and camera domains through a composite loss function, comprising forward/backward photometric losses, cycle-consistency losses, and smoothness regularization.}
    \label{fig: Overview}
\end{figure}

To represent the continuous mapping functions $\mathcal{F}$ and $\mathcal{G}$, we employ two independent coordinate-based Multi-Layer Perceptrons (MLPs) \cite{rumelhart1986learning}, denoted as $\text{Net}_{p2c}$ and $\text{Net}_{c2p}$. Rather than directly regressing absolute coordinates, each network is designed as a residual learner that estimates a local displacement field $\Delta\mathbf{u}$ \cite{he2016deep}. The mapping functions are defined as:
\begin{equation}
    \mathcal{F}(\mathbf{u}_p; \theta_{fwd}) = \mathbf{u}_p + \text{MLP}_{fwd}(\mathbf{u}_p), \quad \mathcal{G}(\mathbf{u}_c; \theta_{bwd}) = \mathbf{u}_c + \text{MLP}_{bwd}(\mathbf{u}_c)
\end{equation}
where $\theta$ represents the learnable parameters. By initializing the output layer weights to zero, the networks initially perform an identity transformation and gradually adapt to the surface-induced geometric shifts.

Each MLP architecture consists of four fully connected layers. Every hidden layer is integrated with Layer Normalization and a LeakyReLU activation function (slope $\alpha=0.2$) to maintain stable gradient flow and prevent activation saturation. To enable end-to-end optimization, we utilize a differentiable bilinear interpolation mechanism called Spatial Transformer \cite{jaderberg2015spatial}. For a given input image $I$, the warped result at a predicted coordinate $\mathbf{u}$ is computed by:
\begin{equation}
    I_{warped}(\mathbf{u}) = \sum_{\mathbf{n} \in \mathcal{N}(\mathbf{u})} I(\mathbf{n}) \prod_{d \in \{x,y\}} (1 - |u_d - n_d|)
\end{equation}
where $\mathcal{N}(\mathbf{u})$ denotes the four-pixel neighborhood of the coordinate. This framework allows the backpropagation of loss signals from the image space directly to the coordinate mapping parameters as shown in Fig.~\ref{fig: Overview}(a).

\subsection{Self-supervised Learning Objective}
The networks are optimized by minimizing a composite loss function $\mathcal{L}_{total}$ that enforces multi-view photometric consistency and geometric constraints:
\begin{equation}
    \mathcal{L}_{total} = w_{fwd}\mathcal{L}_{fwd} + w_{bwd}\mathcal{L}_{bwd} + w_{cyc}\mathcal{L}_{cyc} + w_{sm}\mathcal{L}_{sm} + w_{msk}\mathcal{L}_{msk}
\end{equation}

\subsubsection{Photometric and Structural Consistency}
The primary supervision originates from the pixel-wise and structural alignment between the projected pattern $I_p$ and the captured image $I_c$. The forward loss $\mathcal{L}_{fwd}$ is formulated as:
\begin{equation}
    \mathcal{L}_{fwd} = \alpha \mathcal{L}_{grad}(I_c, \hat{I}_c, M_c) + (1-\alpha) \| (I_c - \hat{I}_c) \odot M_c \|_1
\end{equation}
where $\hat{I}_c = I_p(\mathcal{G}(\mathbf{u}_c))$ is the synthesized camera image and $M_c$ is the binary mask identifying the valid pixels of the projected content in the camera view. The structural term $\mathcal{L}_{grad}$ utilizes a binary Sobel operator $S(\cdot)$ to emphasize edge alignment {clarke1974pattern}:
\begin{equation}
    \mathcal{L}_{grad} = \frac{1}{\sum M} \| (S(I_1) - S(I_2)) \odot M \|_1
\end{equation}
The backward loss $\mathcal{L}_{bwd}$ is defined symmetrically in the projector space.

\subsubsection{Geometric Regularization}
To ensure the bijectivity and physical plausibility of the learned mappings, we introduce a cycle-consistency loss and a smoothness regularizer. The cycle-consistency loss $\mathcal{L}_{cyc}$ constrains the inverse mapping relationship:
\begin{equation}
    \mathcal{L}_{cyc} = \| \mathcal{G}(\mathcal{F}(\mathbf{u}_p)) - \mathbf{u}_p \|_1 + \| (\mathcal{F}(\mathcal{G}(\mathbf{u}_c)) - \mathbf{u}_c) \odot M_c \|_1
\end{equation}

The spatial continuity of the coordinate transformations is enforced via a Total Variation (TV) smoothness loss, $\mathcal{L}_{sm}$ {rudin1992nonlinear}. In physical projection on curved surfaces, the geometric deformation is inherently continuous. Without proper regularization, the high degrees of freedom in MLPs may lead to unphysical grid distortions, such as pixel folding or local discontinuities. By minimizing the squared magnitude of the coordinate gradients, we ensure that the predicted displacement fields remain locally coherent:
\begin{equation}
    \mathcal{L}_{sm} = \sum_{\mathbf{u} \in \Omega_p} \| \nabla \mathcal{F}(\mathbf{u}) \|^2_2 + \sum_{\mathbf{u} \in \Omega_c} \| \nabla \mathcal{G}(\mathbf{u}) \|^2_2
\end{equation}

Furthermore, we introduce a mask consistency loss, $\mathcal{L}_{msk}$, to refine the boundary alignment and account for the Field-of-View (FOV) discrepancy between the hardware components. In practical scenarios, the projected content typically occupies only a sub-region of the camera's captured frame rather than the entire pixel array. To prevent the network from incorrectly mapping projector coordinates to the invalid background pixels in the camera space, we apply a constraint that aligns the warped projector boundary with the detected camera-view mask:
\begin{equation}
    \mathcal{L}_{msk} = \| \text{warp}(M_p, \mathcal{F}) - M_c \|^2_2 + \| \text{warp}(M_c, \mathcal{G}) - M_p \|^2_2
\end{equation}
where $M_p$ is an all-ones identity mask representing the projector's active area, and $M_c$ is the binary mask extracted from the camera image.
This term ensures the coordinate mapping is strictly confined to the effective projection region, thereby improving the precision of the geometric correction.
The full training architecture and self-supervised optimization scheme for \textbf{CSPR-Net} are depicted in Fig.~\ref{fig: Overview}(b).

\section{Results and Discussion}
To systematically evaluate the performance of the  \textbf{CSPR-Net}, we conducted a two-phase validation.
First, a quadric transfer-based ray-tracing simulation was employed to generate synthetic distorted images and analytical ground truth \cite{raskar2004quadric}, which were used to train and quantitatively evaluate the network's reconstruction accuracy.
Second, physical experiments were performed to validate the robustness of the proposed framework under actual operational conditions.
In both simulation and physical trials, the proposed \textbf{CSPR-Net} was benchmarked against a traditional polynomial fitting baseline, consistently demonstrating superior geometric correction performance.

\subsection{Simulation Analysis} \label{Simulation}
\begin{figure}[!ht]
    \centering
    \includegraphics[width=1.0\linewidth]{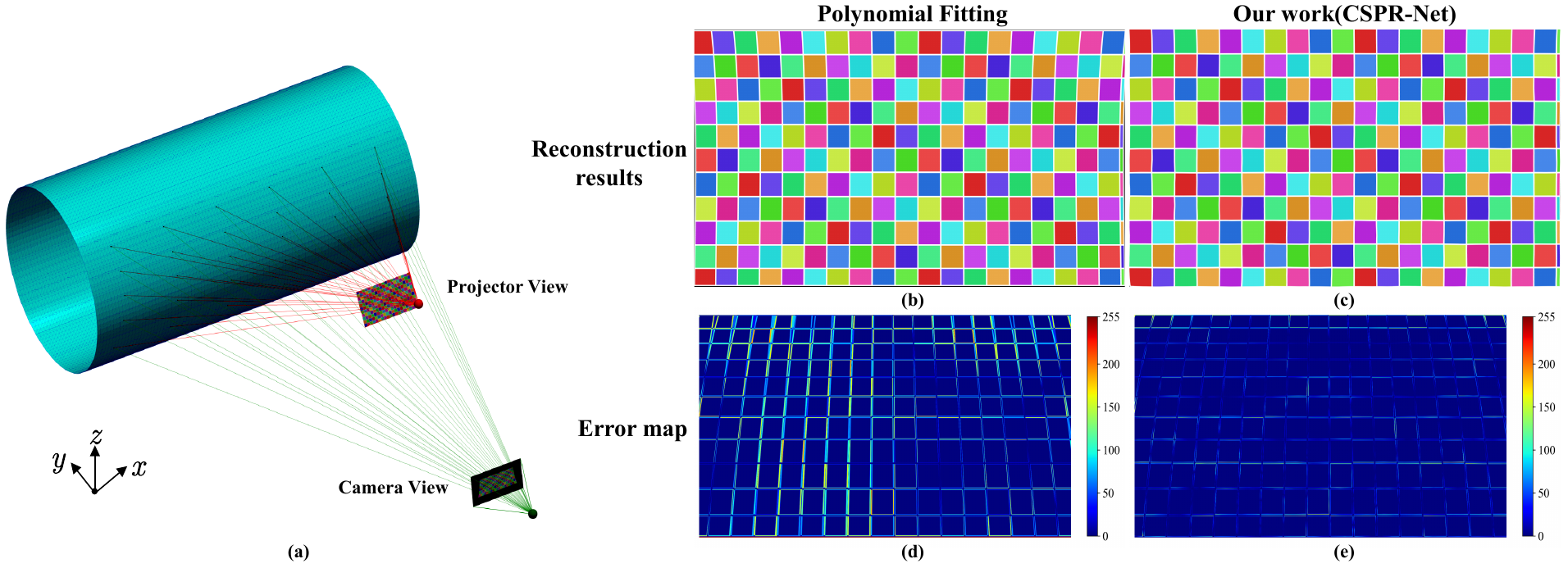}
    \caption{Simulation environment setup and comparative results. (a) The virtual ray-tracing configuration comprising a projector-camera pair and a cylindrical surface; (b)-(c) Comparison of inverse mapping performance, where the proposed \textbf{CSPR-Net} (c) successfully restores a rectilinear grid unlike the polynomial baseline (b) which retains residual distortions; (d)-(e) Pixel-wise error heatmaps relative to the ground truth, demonstrating that our method (e) significantly minimizes reconstruction errors at the boundaries compared to the baseline (d).}
    \label{fig:sim_setup}
\end{figure}

We established a high-fidelity virtual optical bench to isolate geometric non-linearities from environmental noise. A ray-tracing engine was developed to simulate a projector-camera pair in a convergent configuration targeting a vertical cylindrical surface ($R=2.0$). The virtual projector and camera were configured with resolutions of $1920 \times 1080$ and $3072 \times 1728$ respectively. By modeling the surface implicitly as $x^2 + y^2 - R^2 = 0$, we generated the synthesized distorted camera view through forward ray-casting (Projector $\rightarrow$ Surface $\rightarrow$ Camera).
Crucially, the ground truth (GT) pre-warped image was generated via inverse ray-tracing (Camera $\rightarrow$ Surface $\rightarrow$ Projector), providing an absolute reference for geometric accuracy that is often unattainable in physical setups.
The simulation environment and the generated optical paths are visualized in Fig.~\ref{fig:sim_setup}(a).

To mitigate ambient light interference, we employ a specialized calibration pattern featuring spatially varying grid densities and complementary HSL color transitions.
This design captures multi-scale geometric details while inducing high-magnitude boundary gradients.
These strong edge features directly complement our gradient-consistency loss, ensuring robust structural alignment even under complex illumination conditions.

To assess the geometric correction fidelity comprehensively, we employed a multi-dimensional evaluation framework comprising three complementary metrics: Root Mean Square Error (RMSE) to quantify the absolute pixel-wise deviation, Peak Signal-to-Noise Ratio (PSNR) to measure the overall reconstruction quality, and Structural Similarity Index (SSIM) to evaluate perceptual structural alignment \cite{wang2004image}.
Using these metrics, we benchmarked our proposed \textbf{CSPR-Net} against a standard 3rd-degree Polynomial Fitting baseline.
The polynomial approach relies on extracting sparse grid intersections from the distorted view to compute a global parametric mapping, which often lacks the flexibility to model high-frequency local distortions.
In contrast, \textbf{CSPR-Net} is learned directly from raw distorted image pairs via self-supervision, avoiding the errors introduced by explicit feature detection.
Performance was evaluated in three domains: the \textit{Forward Mapping Domain} (mapping original pattern to camera space vs. actual distorted image), the \textit{Inverse Mapping Domain} (mapping distorted view back to projector vs. original pattern), and the \textit{Pre-warped Domain} (generated pattern vs. analytical GT).

As detailed in Table~\ref{tab:sim_results}, the proposed \textbf{CSPR-Net} outperforms the polynomial fitting baseline by a substantial margin across all metrics. In the \textit{Forward Mapping} task, while the polynomial method achieves a reasonable SSIM of 0.9079, it suffers from a relatively high RMSE of 34.02. In contrast, our method reduces the RMSE by approximately 50\% to 17.11, indicating significantly higher pixel-level precision.
The performance gap is even more pronounced in the \textit{Pre-warped Domain}, which directly dictates the quality of the final projection. The polynomial approach struggles to fit high-frequency local non-linearities, resulting in a low PSNR of 14.29 dB and an RMSE of 49.21.
Conversely, \textbf{CSPR-Net} achieves a PSNR of 22.73 dB and an RMSE of 18.60, demonstrating its superior capability to minimize geometric deviations and accurately recover the inverse topology of the curved surface.

Qualitative results further validate these metrics.
As shown in Fig.~\ref{fig:sim_setup}(b) and Fig.~\ref{fig:sim_setup}(c), we visualized the inverse mapping results (transforming the distorted camera view to the projector space). The polynomial result Fig.~\ref{fig:sim_setup}(b) shows visible misalignment compared to the dense and accurate prediction of \textbf{CSPR-Net} Fig.~\ref{fig:sim_setup}(c).
Furthermore, to highlight the precision of the geometric correction, Fig.~\ref{fig:sim_setup}(d) and(e) display the pixel-wise error maps between the generated pre-warped images and the ground truth.
The polynomial method Fig.~\ref{fig:sim_setup}(d) exhibits significant error concentrations near the boundaries where the depth gradient is steep.
In contrast, the error map of \textbf{CSPR-Net} Fig.~\ref{fig:sim_setup}(e) is sparse and low-magnitude, confirming that the network has effectively minimized geometric deviations across the entire curved surface.

\begin{table}[htbp]
    \centering
    \caption{Quantitative comparison of correction performance in simulation}
    \label{tab:sim_results}
    \resizebox{\linewidth}{!}{
        \begin{tabular}{cccccccccc} 
        \specialrule{0.08em}{0em}{0em} 
        \multirow{2}{*}{Method} & 
        \multicolumn{3}{c}{Forward Mapping (vs Distorted)} & 
        \multicolumn{3}{c}{Inverse Mapping (vs Target)} & 
        \multicolumn{3}{c}{Pre-warped (vs GT)} \\ 
        \cmidrule{2-10} 
         & RMSE($\downarrow$ \textsuperscript{a}) & PSNR($\uparrow$) & SSIM($\uparrow$) & RMSE($\downarrow$) & PSNR($\uparrow$) & SSIM($\uparrow$) & RMSE($\downarrow$) & PSNR($\uparrow$) & SSIM($\uparrow$) \\ 
        \specialrule{0.05em}{0em}{0em} 
        
        Polynomial Fitting (Deg=3) & 34.0231 & 17.4953 & 0.9079 & 51.5938 & 13.8789 & 0.7697 & 49.2108 & 14.2896 & 0.7779 \\
        \textbf{Our Work (CSPR-Net)} & \textbf{17.1163} & \textbf{23.4626} & \textbf{0.9603} & \textbf{22.1510} & \textbf{21.2230} & \textbf{0.9227} & \textbf{18.6050} & \textbf{22.7382} & \textbf{0.9282} \\
        \specialrule{0.08em}{0em}{0em} 
        \end{tabular}
    }
    \vspace{-10pt}
    \flushleft\footnotesize \textsuperscript{a}The downward arrow ($\downarrow$) signifies that the metric value is negatively correlated with image quality, and vice versa.
\end{table}

\subsection{Experimental Verification}
To validate the practical robustness of \textbf{CSPR-Net}, we conducted physical experiments using a commercial DLP projector and a CMOS camera. Unlike reconstruction-based methods that demand strict geometric calibration, our framework supports a flexible configuration where the projector and camera can be arbitrarily positioned, provided the camera's field of view fully captures the projection area. 

\begin{figure}[!ht]
    \centering
    \includegraphics[width=1.0\linewidth]{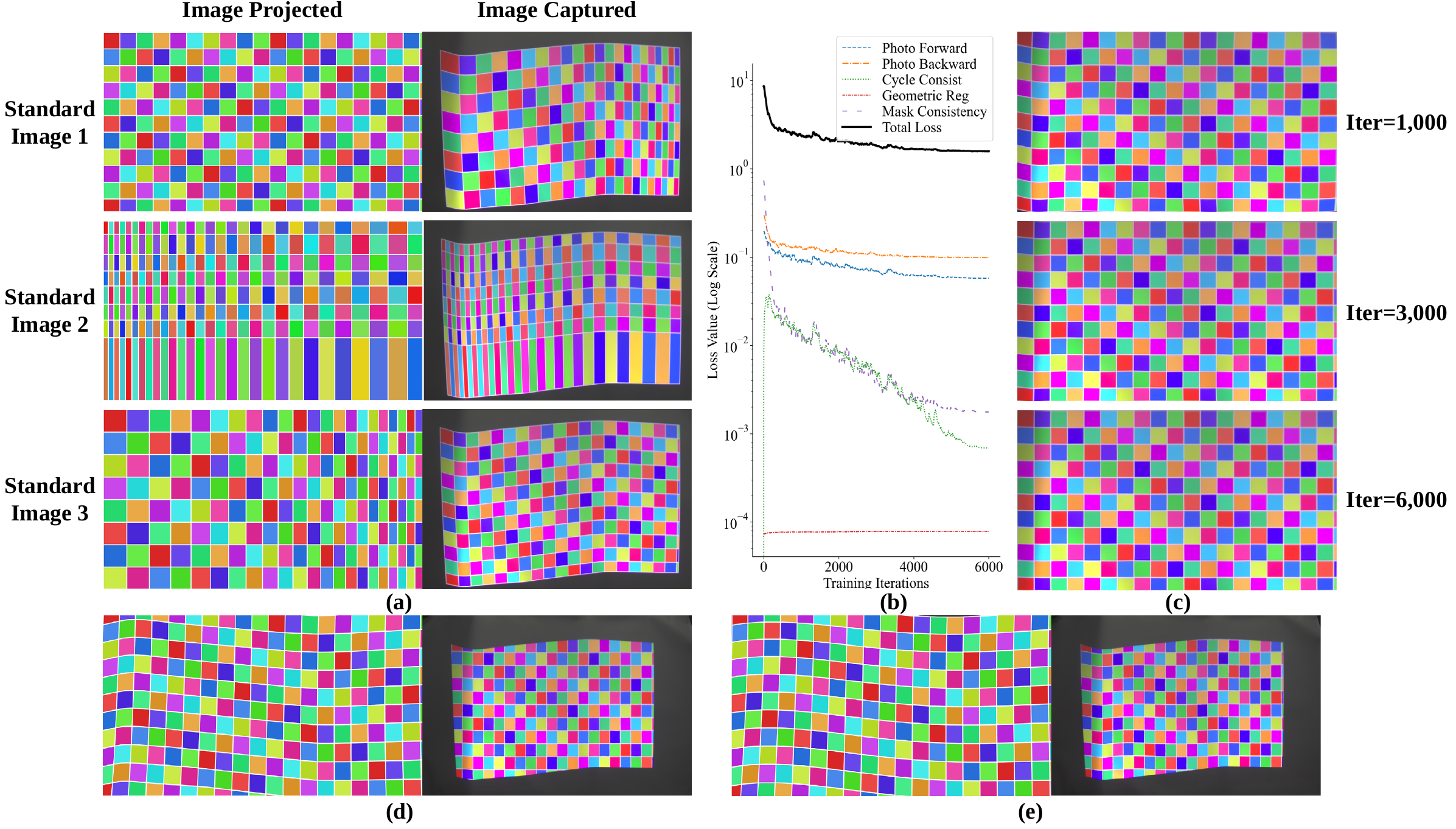}
    \caption{Experimental validation and performance analysis of \textbf{CSPR-Net} on a physical curved surfaces.
    (a) Visualization of the experimental dataset. The data is organized by rows and columns: the first column displays three distinct source calibration patterns featuring spatially varying grid densities and color transitions while the second column presents the corresponding distorted images captured by the camera, which exhibit severe non-linear warping due to the physical surface geometry;
    (b) Convergence analysis of the self-supervised objective. The plot illustrates the stable descent of the composite loss function and its constituent terms, including forward/backward photometric, cycle-consistency, geometric regularization, and mask consistency losses over 6,000 training iterations;
    (c) Visual progression of the inverse geometric mapping during training. Reconstructed projector-space images from distorted camera inputs are shown at $Iter = 1000, 3000,$ and $6000$. As training proceeds, \textbf{CSPR-Net} progressively rectifies the curvilinear distortions to recover the underlying rectilinear grid structure;
    (d) Geometric correction results of the proposed \textbf{CSPR-Net}. This subfigure displays the generated pre-warped image and its corresponding physical display after being projected onto the curved surface, demonstrating high-fidelity rectification;
    (e) Geometric correction results of the polynomial fitting baseline. This subfigure shows the pre-warped image and the associated physical projection result generated by the 3rd-degree polynomial method. Compared to \textbf{CSPR-Net}, the baseline exhibits noticeable residual curvilinear distortion at the image boundaries, highlighting its limitations in accurately modeling complex non-linearities in regions with steep geometric gradients.} 
    \label{fig:exp_results}
\end{figure}

Figure~\ref{fig:exp_results}(a) illustrates the diverse calibration patterns and their corresponding physical captures. 
These source patterns, as detailed in Section~\ref{Simulation}, are specifically engineered with a multi-scale, high-contrast grid structure to ensure robust geometric feature extraction and gradient consistency. 
As observed in the captured samples, the projection undergoes substantial non-linear warping inherent to the complex topography of the physical manifold. 
These distorted image pairs serve as the fundamental input for the subsequent self-supervised rectification process.

The self-supervised learning dynamics are visualized in Fig.~\ref{fig:exp_results}(b) and Fig.~\ref{fig:exp_results}(c). 
Fig.~\ref{fig:exp_results}(b) plots the descent of the constituent loss functions over training iterations on a logarithmic scale, demonstrating stable convergence without ground-truth supervision. 
Coinciding with this numerical convergence, Fig.~\ref{fig:exp_results}(c) visualizes the evolution of the inverse mapping capability ($\text{Net}_{c2p}$). We extracted the reconstructed projector-space images from the distorted camera inputs at $1,000$, $3,000$, and $6,000$ (final) iterations. 
Initially, at iteration $1,000$, the reconstructed image retains significant curvilinear distortion. As training progresses to $3,000$ iterations, the grid structure begins to straighten, and by the final $6,000$ iteration, the network successfully recovers a rectilinear grid that closely matches the original undistorted pattern, confirming that \textbf{CSPR-Net} has effectively learned the inverse topology of the curved surface.

The comparative results are shown in Fig.~\ref{fig:exp_results}(d) and Fig.~\ref{fig:exp_results}(e), where each subfigure displays the pre-warped image on the left and its corresponding physically projected capture on the right. 
In Fig.~\ref{fig:exp_results}(d), the projection rectified by \textbf{CSPR-Net} appears visually flat and undistorted, with the grid lines effectively restored to their rectilinear state. 
Conversely, Fig.~\ref{fig:exp_results}(e) presents the results obtained from the polynomial fitting method. 
While the polynomial approach successfully corrects the macro-level geometry, it fails to resolve local non-linearities, leading to visible residual deformations particularly at the edges and corners of the projected image. 
In these peripheral regions, the projected grid lines are not fully recovered to their rectilinear state, demonstrating the inherent modeling limitations of the parametric baseline when dealing with the complex, high-frequency distortions of the curved surface.

\begin{table}[htbp]
    \centering
    \caption{Quantitative comparison of correction performance in physical experiments}
    \label{tab:exp_quantitative}
    
    \resizebox{\linewidth}{!}{%
        \begin{tabular}{ccccccccc} 
        \specialrule{0.08em}{0em}{0em}
        \multirow{2}{*}{Method} & 
        \multicolumn{2}{c}{End-to-End Fidelity (SSIM $\uparrow$ \textsuperscript{a})} & 
        \multicolumn{3}{c}{Forward Mapping (vs Distorted)} & 
        \multicolumn{3}{c}{Inverse Mapping (vs Target)} \\ 
        \cmidrule{2-9} 
         & Before & After & RMSE($\downarrow$) & PSNR($\uparrow$) & SSIM($\uparrow$) & RMSE($\downarrow$) & PSNR($\uparrow$) & SSIM($\uparrow$) \\ 
        \specialrule{0.05em}{0em}{0em}
        
        Polynomial Fitting (Deg=3) & 0.5384 & 0.6211 & 41.9776 & 15.6705 & 0.7112 & 43.7622 & 15.2692 & 0.6902 \\
        \textbf{Our Work (CSPR-Net)} & \textbf{0.5384} & \textbf{0.6501} & \textbf{37.2289} & \textbf{16.7132} & \textbf{0.7383} & \textbf{38.2367} & \textbf{16.4812} & \textbf{0.7276} \\
        \specialrule{0.08em}{0em}{0em}
        \end{tabular}%
    }
    \vspace{-10pt}
    \flushleft\footnotesize \textsuperscript{a}The upward arrow ($\uparrow$) signifies that the metric value is positively correlated with image quality, and vice versa.
\end{table}

Finally, we performed a comprehensive quantitative evaluation of the mapping accuracy and correction efficacy in the physical experimental scenarios, as summarized in Table~\ref{tab:exp_quantitative}. 
First, the \textit{End-to-End Fidelity} metrics provide a rigorous quantitative foundation for the visual rectification results presented in Fig.~\ref{fig:exp_results}(a). 
Without any compensation, the direct projection on the curved manifold suffers from a low SSIM of 0.5384 due to severe curvilinear warping. 
By applying the pre-warped field generated by our framework, the SSIM is significantly elevated to 0.6501, which substantiates that \textbf{CSPR-Net} effectively mitigates geometric distortions and successfully restores the rectilinear topology of the projected content.

Beyond proving general efficacy, the experimental data demonstrates that \textbf{CSPR-Net} outperforms the 3rd-degree polynomial baseline by a substantial margin across all three evaluation metrics: RMSE, PSNR, and SSIM. 
As detailed in the \textit{Forward} and \textit{Inverse Mapping} domains, our method achieves consistently lower pixel-wise error (RMSE) alongside higher reconstruction quality (PSNR) and structural alignment (SSIM). 
Specifically, in the forward mapping task, our approach reduces the RMSE from 41.98 to 37.23 and improves the PSNR to 16.71 dB, indicating a much higher precision in modeling complex, high-frequency non-linearities compared to the global parametric fitting. 
The simultaneous improvement across this multi-dimensional framework confirms the superior robustness and fidelity of the proposed self-supervised neural architecture in physical environments.

\begin{figure}[!ht]
    \centering
    \includegraphics[width=1.0\linewidth]{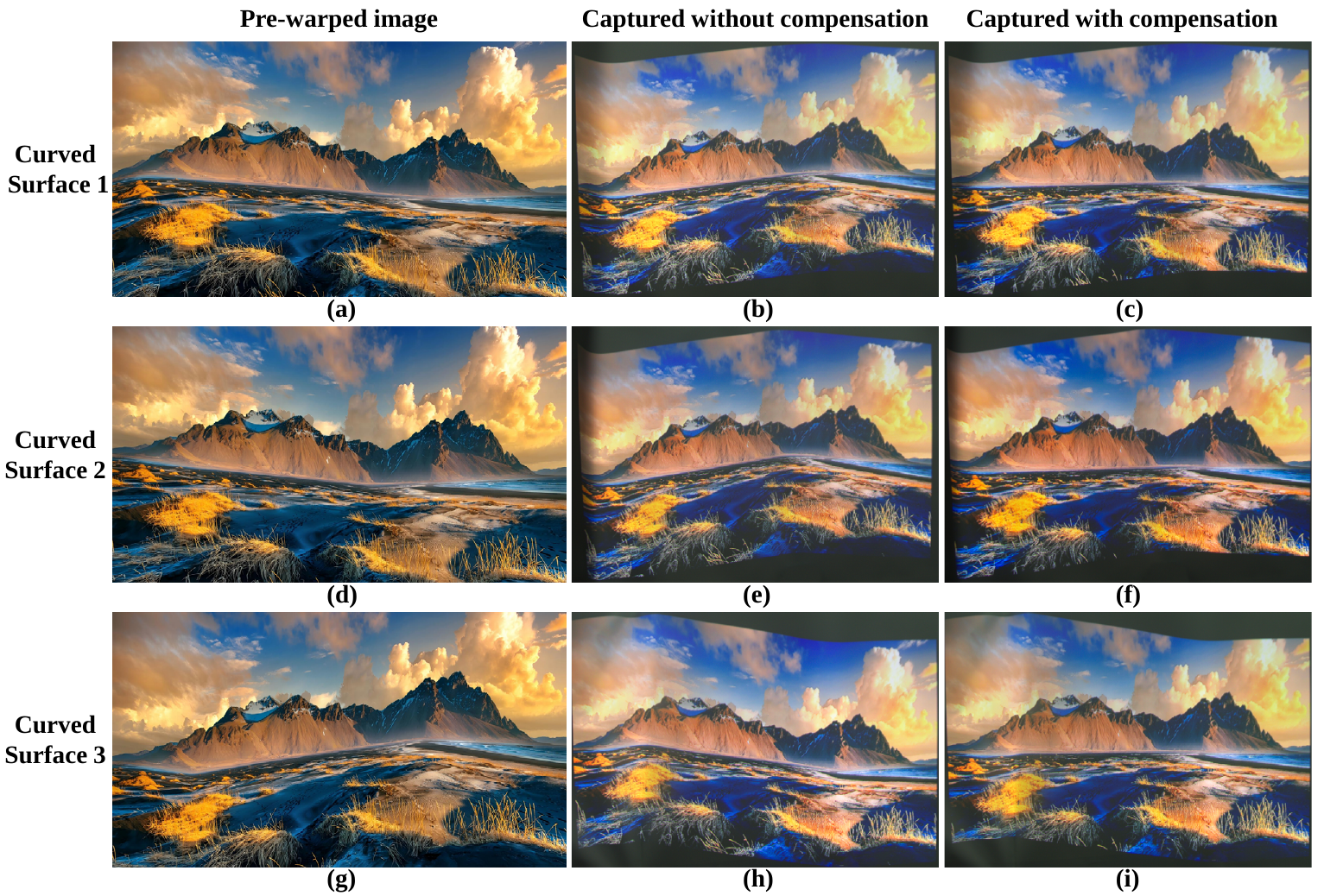}
    \caption{Qualitative evaluation of \textbf{CSPR-Net} on three distinct curved surfaces using an unseen scenic test image. 
    From top to bottom, the rows represent image sets captured on W, V, and M-shaped surfaces respectively. 
    The columns represent the stages of the rectification process: 
    the first column (a, d, g) displays the pre-warped images generated by our framework; 
    the second column (b, e, h) shows the direct projection results without any geometric compensation; 
    and the third column (c, f, i) captures the final rectified projections.
    The results demonstrate that \textbf{CSPR-Net} consistently achieves high-fidelity, rectilinear displays across diverse complex topographies.}
    \label{fig:multi_surface_comparison}
\end{figure}

\begin{table}[!ht]
    \centering
    \caption{Quantitative comparison across distinct surface geometries}
    \label{tab:multi_surface_comparison}
    
    \resizebox{\linewidth}{!}{
        \begin{tabular}{ccccccc} 
        \specialrule{0.08em}{0em}{0.5ex}
        \multirow{2}{*}{\textbf{Surface Sequence}} & 
        \multicolumn{3}{c}{\textbf{Direct Projection}} & 
        \multicolumn{3}{c}{\textbf{Rectified Projection}} \\ 
        \cmidrule{2-4} \cmidrule{5-7} 
         & RMSE ($\downarrow$ \textsuperscript{a}) & PSNR ($\uparrow$) & SSIM ($\uparrow$) & RMSE ($\downarrow$) & PSNR ($\uparrow$) & SSIM ($\uparrow$) \\ 
        \specialrule{0.05em}{0.5ex}{0.5ex}
        
        Surface 1 & 47.2146 & 14.6493 & 0.4021 & \textbf{38.1339} & \textbf{16.5046} & \textbf{0.4295} \\
        Surface 2 & 50.6026 & 14.0473 & 0.4208 & \textbf{41.1570} & \textbf{15.8419} & \textbf{0.4645} \\
        Surface 3 & 52.2143 & 13.7750 & 0.4097 & \textbf{39.9156} & \textbf{16.1079} & \textbf{0.4502} \\
        \specialrule{0.08em}{0.5ex}{0em}
        \end{tabular}
    }
    \vspace{-10pt}
    \flushleft\footnotesize \textsuperscript{a}The down arrow ($\downarrow$) signifies that the metric value is negatively correlated with image quality, and vice versa.
\end{table}

To further evaluate the versatility and robust generalization capability of \textbf{CSPR-Net}, we extended the testing phase beyond the structured calibration patterns used during training. 
While the network was optimized using regular, feature-rich grid patterns , we selected a high-resolution scenic landscape image characterized by a pseudo-random distribution of pixel intensities and complex color transitions as an unseen test sample. 
This validation was conducted across three complex non-planar manifolds with distinct geometric profiles: Surface 1 features a W-shaped geometry (dual concave regions flanking a central convexity); Surface 2 presents a V-shaped asymmetric slope; and Surface 3 is configured as an M-shaped manifold.

The qualitative results are visualized in Fig.~\ref{fig:multi_surface_comparison}, where each row demonstrates the model's performance on these different challenging topographies. 
As shown in the second column, direct projection of the original image suffers from severe and varying non-linear warping inherent to each specific setup. 
Conversely, by applying the inverse distortion fields encoded in the pre-warped images (first column), \textbf{CSPR-Net} successfully compensates for the physical deformations, resulting in geometrically consistent and visually rectilinear displays (third column). 

This visual rectification is further supported by the quantitative data summarized in Table~\ref{tab:multi_surface_comparison}, which benchmarks the \textit{Direct Projection} against our \textit{Rectified Projection}. 
Despite the complexity of the W, V, and M surface profiles, our framework achieves significant improvements across all three RMSE, PSNR, and SSIM metrics. 
Specifically, the average RMSE is reduced by approximately 20\%, while the SSIM scores consistently increase, confirming that the self-supervised framework can robustly adapt to diverse surface geometries and deliver high-fidelity corrections for arbitrary, complex visual content.

\section{Conclusion}
In this paper, we presented \textbf{CSPR-Net}, a self-supervised framework that redefines geometric correction for non-planar projection. 
By treating distortion rectification as a cycle-consistent learning problem via implicit coordinate representations, our approach effectively overcomes the trade-off between calibration complexity and correction precision inherent in traditional parametric models.
Macro-level analysis indicates that this data-driven paradigm offers superior flexibility, autonomously adapting to arbitrary surface topologies without the need for physical measurements or explicit ground-truth deformation fields.
Experimental results, both in high-fidelity simulations and physical setups, confirm that \textbf{CSPR-Net} consistently outperforms polynomial fitting baselines, delivering significantly higher structural similarity and visually coherent pre-warped projections.
While this work addresses geometric non-linearities, practical projection systems also face photometric and optical challenges.
Future work will focus on extending this neural architecture to a holistic compensation system, integrating modules for radiometric color correction and defocus blur restoration to ensure high-fidelity display in complex optical environments.

\begin{backmatter}

\bmsection{Fundings}
This study was supported by the Zhejiang Provincial Natural Science Foundation of China (LR25F050002) and National Natural 
Science Foundation of China (82371709).

\bmsection{Disclosures}
The authors declare no conflicts of interest.

\bmsection{Data Availability Statement}
The image datasets and experimental results utilized in this study were independently collected and generated by the authors. All relevant data supporting the findings are presented in full through the figures and tables included within this manuscript. The source code will be publicly accessible upon acceptance of this manuscript. Additional relevant information is available by reasonable request from the corresponding author X.H.
\end{backmatter}

\bibliography{sample}

\begin{thebibliography}{10}
\newcommand{\enquote}[1]{``#1''}

\bibitem{grundhofer2018recent}
A.~Grundh{\"o}fer and D.~Iwai, \enquote{Recent advances in projection mapping algorithms, hardware and applications,} in \emph{Computer graphics forum,}  vol.~37 (Wiley Online Library, 2018), pp. 653--675.

\bibitem{bimber2005spatial}
O.~Bimber and R.~Raskar, \emph{Spatial augmented reality: merging real and virtual worlds} (CRC press, 2005).

\bibitem{park2016defocus}
J.~Park and B.-U. Lee, \enquote{Defocus and geometric distortion correction for projected images on a curved surface,} {\protect\JournalTitle{Applied optics}} \textbf{55}, 896--902 (2016).

\bibitem{ibrahim2024spatially}
M.~T. Ibrahim, \enquote{Spatially augmented reality on dynamic, deformable surfaces and its applications,} Ph.D. thesis, University of California, Irvine (2024).

\bibitem{tehrani2019automated}
M.~A. Tehrani, M.~Gopi, and A.~Majumder, \enquote{Automated geometric registration for multi-projector displays on arbitrary 3d shapes using uncalibrated devices,} {\protect\JournalTitle{IEEE transactions on visualization and computer graphics}} \textbf{27}, 2265--2279 (2019).

\bibitem{moreno2012simple}
D.~Moreno and G.~Taubin, \enquote{Simple, accurate, and robust projector-camera calibration,} in \emph{2012 Second International Conference on 3D Imaging, Modeling, Processing, Visualization \& Transmission,}  (IEEE, 2012), pp. 464--471.

\bibitem{jordan2010projector}
S.~Jordan and M.~Greenspan, \enquote{Projector optical distortion calibration using gray code patterns,} in \emph{2010 IEEE Computer Society Conference on Computer Vision and Pattern Recognition-Workshops,}  (IEEE, 2010), pp. 72--79.

\bibitem{kio2016distortion}
O.~G. Kio, \enquote{Distortion correction for non-planar deformable projection displays through homography shaping and projected image warping,} Ph.D. thesis, University of Central Lancashire (2016).

\bibitem{liu2015accurate}
M.~Liu, C.~Sun, S.~Huang, and Z.~Zhang, \enquote{An accurate projector calibration method based on polynomial distortion representation,} {\protect\JournalTitle{Sensors}} \textbf{15}, 26567--26582 (2015).

\bibitem{fitzgibbon2001simultaneous}
A.~W. Fitzgibbon, \enquote{Simultaneous linear estimation of multiple view geometry and lens distortion,} in \emph{Proceedings of the 2001 IEEE Computer Society Conference on Computer Vision and Pattern Recognition. CVPR 2001,}  vol.~1 (IEEE, 2001), pp. I--I.

\bibitem{xu2007robust}
Y.~Xu and D.~G. Aliaga, \enquote{Robust pixel classification for 3d modeling with structured light,} in \emph{Proceedings of Graphics Interface 2007,}  (2007), pp. 233--240.

\bibitem{manevarthe2018geometric}
B.~Manevarthe and R.~Kalpathi, \enquote{Geometric correction for projection on non planar surfaces using point clouds,} in \emph{Proceedings of the 12th International Conference on Distributed Smart Cameras,}  (2018), pp. 1--6.

\bibitem{lecun2015deep}
Y.~LeCun, Y.~Bengio, and G.~Hinton, \enquote{Deep learning,} {\protect\JournalTitle{nature}} \textbf{521}, 436--444 (2015).

\bibitem{barbastathis2019use}
G.~Barbastathis, A.~Ozcan, and G.~Situ, \enquote{On the use of deep learning for computational imaging,} {\protect\JournalTitle{Optica}} \textbf{6}, 921--943 (2019).

\bibitem{krizhevsky2012imagenet}
A.~Krizhevsky, I.~Sutskever, and G.~E. Hinton, \enquote{Imagenet classification with deep convolutional neural networks,} {\protect\JournalTitle{Advances in neural information processing systems}} \textbf{25} (2012).

\bibitem{li2019blind}
X.~Li, B.~Zhang, P.~V. Sander, and J.~Liao, \enquote{Blind geometric distortion correction on images through deep learning,} in \emph{Proceedings of the IEEE/CVF conference on computer vision and pattern recognition,}  (2019), pp. 4855--4864.

\bibitem{jin2023perspective}
L.~Jin, J.~Zhang, Y.~Hold-Geoffroy, \emph{et~al.}, \enquote{Perspective fields for single image camera calibration,} in \emph{Proceedings of the IEEE/CVF Conference on Computer Vision and Pattern Recognition,}  (2023), pp. 17307--17316.

\bibitem{wang2024implicit}
Z.~Wang \emph{et~al.}, \enquote{Where do we stand with implicit neural representations? a technical and performance survey,} {\protect\JournalTitle{arXiv preprint arXiv:2411.03688}}  (2024).

\bibitem{huang2019compennet}
B.~Huang and H.~Ling, \enquote{Compennet++: End-to-end full projector compensation,} in \emph{Proceedings of the IEEE/CVF International Conference on Computer Vision,}  (2019), pp. 7164--7173.

\bibitem{isola2017image}
P.~Isola, J.-Y. Zhu, T.~Zhou, and A.~A. Efros, \enquote{Image-to-image translation with conditional adversarial networks,} in \emph{Proceedings of the IEEE conference on computer vision and pattern recognition,}  (2017), pp. 1125--1134.

\bibitem{zhu2017unpaired}
J.-Y. Zhu, T.~Park, P.~Isola, and A.~A. Efros, \enquote{Unpaired image-to-image translation using cycle-consistent adversarial networks,} in \emph{Proceedings of the IEEE international conference on computer vision,}  (2017), pp. 2223--2232.

\bibitem{rumelhart1986learning}
D.~E. Rumelhart, G.~E. Hinton, and R.~J. Williams, \enquote{Learning representations by back-propagating errors,} {\protect\JournalTitle{nature}} \textbf{323}, 533--536 (1986).

\bibitem{he2016deep}
K.~He, X.~Zhang, S.~Ren, and J.~Sun, \enquote{Deep residual learning for image recognition,} in \emph{Proceedings of the IEEE conference on computer vision and pattern recognition,}  (2016), pp. 770--778.

\bibitem{jaderberg2015spatial}
M.~Jaderberg, K.~Simonyan, A.~Zisserman \emph{et~al.}, \enquote{Spatial transformer networks,} {\protect\JournalTitle{Advances in neural information processing systems}} \textbf{28} (2015).

\bibitem{raskar2004quadric}
R.~Raskar, J.~Van~Baar, T.~Willwacher, and S.~Rao, \enquote{Quadric transfer for immersive curved screen displays,} in \emph{Computer Graphics Forum,}  vol.~23 (Wiley Online Library, 2004), pp. 451--460.

\bibitem{wang2004image}
Z.~Wang, A.~C. Bovik, H.~R. Sheikh, and E.~P. Simoncelli, \enquote{Image quality assessment: from error visibility to structural similarity,} {\protect\JournalTitle{IEEE transactions on image processing}} \textbf{13}, 600--612 (2004).

\end{thebibliography}

\end{document}